\documentclass[pra]{revtex4}
\usepackage{amssymb}
\usepackage{amsmath}
\usepackage{graphicx}
\usepackage{dcolumn}
\usepackage[center]{subfigure}
\usepackage{array}
\usepackage{booktabs}
\usepackage{multirow}
\usepackage{float}
\usepackage{color}
\usepackage{ulem}
\usepackage{booktabs}

\begin{document}

\title{Semidiscrete optical vortex droplets in quasi-phase-matched photonic crystals}
\author{Xiaoxi Xu$^{1}$}
\author{Feiyan Zhao$^{1}$}
\author{Jiayao Huang$^{2}$}
\author{Hexiang He$^{1}$}
\author{Li Zhang$^{1}$}
\author{Zhaopin Chen$^{3}$}
\author{Zhongquan Nie$^{4}$}
\author{Boris A. Malomed$^{5,6}$}
\author{Yongyao Li$^{1,7}$}
\email{yongyaoli@gmail.com}
\affiliation{$^{1}$ School of Physics and Optoelectronic Engineering, Foshan University,
Foshan 528000, China \\
$^{2}$School of Electronic and Computer Engineering, Peking University,
Shenzhen 518055, China \\
$^{3}$Physics Department and Solid-State Institute, Technion, Haifa 32000,
Israel\\
$^{4}$Key Lab of Advanced Transducers and Intelligent Control System,
Ministry of Education and Shanxi Province, College of Electronic Information
and Optical Engineering, Taiyuan University of Technology, Taiyuan 030024,
China \\
$^{5}$ Department of Physical Electronics, School of Electrical Engineering,
Faculty of Engineering, Tel Aviv University, Tel Aviv 69978, Israel \\
$^{6}$ Instituto de Alta Investigaci\'{o}n, Universidad de Tarapac\'{a},
Casilla 7D, Arica, Chile \\
$^{7}$ Guangdong-Hong Kong-Macao Joint Laboratory for Intelligent Micro-Nano
Optoelectronic Technology, Foshan University, Foshan 528000, China}
\date{\today}

\begin{abstract}
A new scheme for producing semidiscrete self-trapped vortices
(\textquotedblleft swirling photon droplets\textquotedblright ) in photonic
crystals with competing quadratic ($\chi ^{(2)}$) and self-defocusing cubic (%
$\chi ^{(3)}$) nonlinearities is proposed. The photonic crystal is designed
with a striped structure, in the form of spatially periodic modulation of
the $\chi ^{(2)}$ susceptibility, which is imposed by the
quasi-phase-matching technique. Unlike previous realizations of semidiscrete
optical modes in composite media, built as combinations of continuous and
arrayed discrete waveguides, the semidiscrete vortex \textquotedblleft
droplets\textquotedblright\ are produced here in the fully continuous
medium. This work reveals that the system supports two types of semidiscrete
vortex droplets, \textit{viz}., onsite- and intersite-centered ones, which
feature, respectively, odd and even numbers of stripes, $\mathcal{N}$.
Stability areas for the states with different values of $\mathcal{N}$ are
identified in the system's parameter space. Some stability areas overlap
with each other, giving rise to the multistability of states with different $%
\mathcal{N}$. The coexisting states are mutually degenerate, featuring equal
values of the Hamiltonian and propagation constant. An experimental scheme
to realize the droplets is outlined, suggesting new possibilities for the
long-distance transmission of nontrivial vortex beams in nonlinear media.

\end{abstract}

\maketitle

\section{Introduction}

Semidiscrete vortex quantum droplets, a new type of vortices, were initially
predicted in binary Bose-Einstein condensates trapped in an array of
tunnel-coupled quasi-1D potential wells \cite{Zhang2019}.
Unlike vortex modes in fully continuous or fully discrete systems \cite{Malomed2001,Lederer2008,Malomed2019,Malomed2020,Zhao2022,Xu2023},
these are stripe-shaped localized states, which are continuous in one
direction and discrete in the perpendicular one, and do not exhibit
rotational symmetry. It is well known that the stability of self-trapped
vortex modes in two-dimensional (2D) and three-dimensional (3D) geometries
is a challenging problem because the self-attractive nonlinearity gives rise
to strong splitting instability of vortex rings and tori, even if the
collapse instability that affects fundamental (zero-vorticity) solitons in
the same media may be suppressed \cite{Berge1998,Trapani2000,Mihalache2004CQ,Sulem2007,Fibich2015,Malomed2022}. Due to the competition between the mean-field (MF) and beyond-MF effects in the bosonic condensate \cite{Petrov2015,Petrov2016,Schmitt2016,Cabrera2018,Kartashov2018,Li2018,Luo2021,Lin2021,Zheng2021,Li2023}%
, semidiscrete vortex quantum droplets may maintain stability in this
setting against the azimuthal (splitting) perturbations. In the field of
nonlinear optics, somewhat similar objects in the form of \textquotedblleft
photon droplets" were experimentally demonstrated in optical media with
nonlocal (thermal) nonlinearity \cite{Wilson2018,Westerberg2018}. Actually, the competition between
different nonlinear terms is a common effect \cite{Buryak1995,Mihalache2004,Desyatnikov2005,Wu2013,DeSalvo1992,Chen2004,Bosshard1995}%
, which occurs in the propagation of high-power laser beams in various media %
\cite{Bosshard1995,Cid2017}. Optical semidiscrete vortex
droplets can be maintained by the balance between the competing
nonlinearities. In particular, stable self-bound semidiscrete vortex modes
in the spatial domain were predicted in coupled planar waveguides with the
cubic-quintic nonlinearity \cite{Xu2021}. Similarly,
self-bound spatiotemporal vortex modes can be predicted in coupled arrays of
nonlinear fibers \cite{Leblond}.

Recently, patterned quasi-phase matched (QPM) nonlinear photonic crystals in
the 3D space have been produced by means of the thermoelectric field
polarization \cite{Xu2018}, laser erasing %
\cite{Wei2018}, and femtosecond laser poling technique %
\cite{Shanliu2023}, which provides more possibilities for
the creation of vortex states. The QPM\ technique has developed to a
well-known method for achieving accurate phase matching in $\chi ^{(2)}$
crystals for the nonlinear frequency conversion \cite{Zhu1997,Clausen1997,Bang1999,Corney2001,Arie2010,Thoma2013,Hu2013,Karnieli18,Karnieli19,Lin2019,Li2020}
and nonlinear beam shaping \cite{Bloch2012,Asia2012,Asia2013,Mills2015,XuOE2018,Wei2019,Imbrock2020,Chen2021}
in different dimensions. Very recently, stable vortex solitons were
predicted in 3D QPM photonic crystals \cite{Zhao2023}. The
structure of the vortex solitons can be engineered by fixing different
phase-matching conditions in different cells of the photonic crystals, thus
inducing effective discreteness in this 3D optical medium. This technique
offers a possibility of building semidiscrete vortex modes in the
QPM-structured bulk photonic crystals. It is relevant to mention that
effective 2D (but not 3D) discrete waveguiding structures for optical beams
with the extraordinary polarization can be induced by means of a different
technique in photorefractive materials illuminated by interfering beams with
the ordinary polarization\textsuperscript{\cite{Segev1}}. Such virtual
photonic lattices were used for the creation of quasi-discrete 2D vortex
solitons \cite{Segev2,Kivshar,Denz}.

\begin{figure}[th]
\centering
{\includegraphics[scale=0.4]{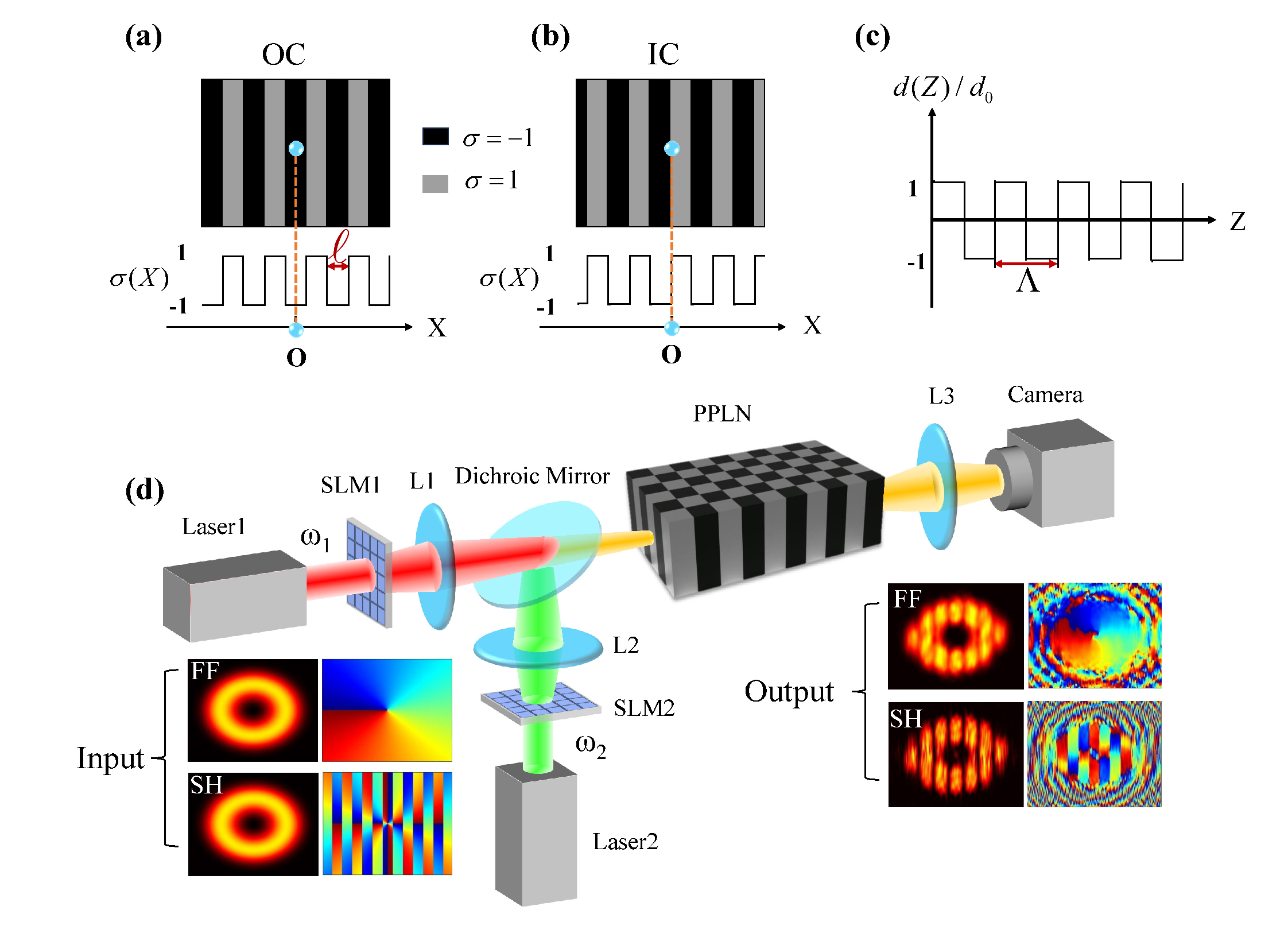}}
\caption{(Color online) (a,b) The structures corresponding to OC and IC
modulations, which are defined as per Eq. (\ref{ICOC}), $\ell $ and $\mathbf{O}$ representing the stripe's width and center of the modulation pattern, respectively. The black and gray blocks represent $\sigma=-1$ and $+1$, respectively. (c) The periodic modulation along the $Z$ axis defined as per Eq. (\ref{Dz1}), $\Lambda $ being the period of the
longitudinal modulation. (d) A schematic of the experimental setup for the
creation of the semidiscrete vortex droplets: L1, L2, L3 -- lenses, SLM1,
SLM2 -- spatial light modulators, PPLN -- the periodically polarized lithium
niobate crystal, FF -- the fundamental frequency, SH -- the second harmonic.
Two bottom plots display the input and output intensity and phase patterns
of the FF and SH beams.}
\label{stripefig}
\end{figure}

In this paper, we propose a scenario for the creation of semidiscrete vortex
optical droplets in 3D photonic crystals with quadratic ($\chi ^{(2)}$) and
defocusing cubic ($\chi ^{(3)}$) nonlinearities and a striped QPM-induced
spatial structure, as shown in Fig. \ref{stripefig}(a,b,c). The $\chi ^{(3)}$
nonlinearity is able to compete with the $\chi ^{(2)}$ interaction if
intensities of the light fields reach the level of several GW/cm$^{2}$ %
\cite{Bosshard1995}, making it possible to create
self-bound photon droplets. We demonstrate that the phase-matching
condition, adjusted to the striped structure, induces effective discreteness
between adjacent stripes, which is necessary for the design of semidiscrete
states. The balance of the competition between quadratic and cubic
nonlinearity allows the self-bound vortex modes to maintain their stability.

The subsequent material is arranged as follows. The model is introduced in
Section 2. Numerical results and estimates for the experimental setup are
presented in Section 3. The paper is concluded by Section 4.

\section{The model}

The paraxial propagation of light beams through the 3D QPM photonic crystals
with the competing $\chi ^{(2)}$ and $\chi ^{(3)}$ nonlinearities is
governed by coupled equations for the slowly varying fundamental frequency
(FF) and second harmonic (SH) amplitudes, $A_{1}$ and $A_{2}$:

\begin{eqnarray}
&&i\partial _{Z}A_{1}=-\frac{1}{2k_{1}}\nabla ^{2}A_{1}-\frac{2d(X,Z)\omega
_{1}}{cn_{1}}A_{1}^{\ast }A_{2}e^{-i\Delta k_{0}Z}+\frac{3\chi ^{(3)}\omega
_{1}}{2cn_{1}}(|A_{1}|^{2}+2|A_{2}|^{2})A_{1},  \notag \\
&&i\partial _{Z}A_{2}=-\frac{1}{2k_{2}}\nabla ^{2}A_{2}-\frac{d(X,Z)\omega
_{2}}{cn_{2}}A_{1}^{2}e^{i\Delta k_{0}Z}+\frac{3\chi ^{(3)}\omega _{2}}{%
2cn_{2}}(|A_{2}|^{2}+2|A_{1}|^{2})A_{2},  \label{dynamiceq2}
\end{eqnarray}%
where $\nabla ^{2}=\partial _{X}^{2}+\partial _{Y}^{2}$ is the
paraxial-diffraction operator, $c$ is the speed of light in vacuum, while $%
n_{1,2}$, $\omega _{1,2}$ ($\omega _{2}=2\omega _{1}$), and $k_{1,2}$ are,
respectively, the refractive indices, carrier frequencies, and wavenumbers
of the FF and SH components, and $\Delta k_{0}=2k_{1}-k_{2}$ is the
phase-velocity mismatch. $\chi ^{(3)}>0$ is the third-order susceptibility,
which accounts for the cubic self-defocusing. The local modulation of the
second-order susceptibility $\chi ^{(2)}$ is determined by
\begin{equation}
d(X,Z)=\sigma (X)d(Z),  \label{dsd}
\end{equation}%
where $\sigma (X)$ is the transverse striped OC (onsite-centered) or IC
(intersite-centered) modulation pattern:
\begin{equation}
\sigma (X)=%
\begin{cases}
-\mathrm{sgn}\left[ \cos (\pi X/\ell )\right] & \mathrm{OC}, \\
-\mathrm{sgn}\left[ \sin (\pi X/\ell )\right] & \mathrm{IC},%
\end{cases}
\label{ICOC}
\end{equation}%
$\ell $ being the width of a stripe. The OC and IC patterns correspond to
the pivot of the vortex beam located, respectively, at the center of a
stripe or at the border between two stripes, see Fig. \ref{stripefig}(a,b).
Further, the factor accounting in Eq. (\ref{dsd}) for the modulation in the $%
Z$ direction, with amplitude $d_{0}$ and period $\Lambda $, is %
\cite{Suchowski,Karnieli2018,Karnieli2022}
\begin{equation}
d(Z)=d_{0}\mathrm{sgn}\left[ \cos (2\pi Z/\Lambda )\right] \equiv
d_{0}\sum_{m\neq 0}\left( \frac{2}{m\pi }\right) \sin \left( \frac{m\pi }{2}%
\right) \exp \left( i\frac{2\pi m}{\Lambda }Z\right) ,  \label{Dz1}
\end{equation}%
see Fig. \ref{stripefig}(c). Actually, only the terms with $m=\pm 1$ are
kept in Eq. (\ref{Dz1}), as they play the dominant role in the QPM effect.
Thus, $m=1$ and $-1$ relate to the FF and SH components, respectively.

By means of rescaling \cite{Yongyao2020,feiyan2021}
\begin{eqnarray}
&&\zeta =\left( \frac{n_{1}}{\omega _{1}}+\frac{n_{2}}{\omega _{2}}\right)
,\quad z_{d}^{-1}=\frac{2}{c\pi }\frac{d_{0}^{2}}{\chi ^{(3)}}\left( \frac{%
\omega _{1}^{2}\omega _{2}}{n_{1}^{2}n_{2}}\zeta \right) ^{\frac{1}{2}},
\notag \\
&&u_{p}=\frac{\chi ^{(3)}}{d_{0}}\sqrt{\frac{n_{p}}{\omega _{p}\zeta }}%
A_{p}\exp \left[ i(\Delta k_{0}-2\pi /\Lambda )Z\right] ,\quad p=1,2,  \notag
\\
&&z=Z/z_{d},\quad x=\sqrt{k_{1}/z_{d}}X,\quad y=\sqrt{k_{1}/z_{d}}Y,\quad
\Omega =\left( \Delta k_{0}-2\pi /\Lambda \right) z_{d},  \notag \\
&&g_{11}=\frac{3\pi }{4}\sqrt{\frac{\omega _{1}^{2}n_{2}}{n_{1}^{2}\omega
_{2}}\zeta },\quad g_{22}=\frac{3\pi }{4}\sqrt{\frac{\omega ^{3}n_{1}^{2}}{%
n_{2}^{3}\omega _{1}^{2}}\zeta }\quad g_{12}=\frac{3\pi }{2}\sqrt{\frac{%
\omega _{2}}{n_{2}}\zeta },  \label{units}
\end{eqnarray}%
Eqs. (\ref{dynamiceq2}), which keep, as said above, the terms with $m=\pm 1$
in Eq. (\ref{Dz1}), are simplified to the form of
\begin{eqnarray}
&&i\partial _{z}u_{1}=-\frac{1}{2}\nabla ^{\prime 2}u_{1}-\Omega
u_{1}-2\sigma (x)u_{1}^{\ast }u_{2}+\left(
g_{11}|u_{1}|^{2}+g_{12}|u_{2}|^{2}\right) u_{1},  \label{GPeq1} \\
&&i\partial _{z}u_{2}=-\frac{1}{2\eta }\nabla ^{\prime 2}u_{2}-\Omega
u_{2}-\sigma (x)u_{1}^{2}+\left( g_{22}|u_{2}|^{2}+g_{12}|u_{1}|^{2}\right)
u_{2},  \label{GPeq2}
\end{eqnarray}%
where $\nabla ^{^{\prime }2}=\partial _{xx}+\partial _{yy}$ and $\eta
=k_{2}/k_{1}$. Neglecting the slight difference in the FF and SH refractive
indices, i.e., setting $n_{1}=n_{2}$, results in coefficients $g_{11}=3\pi
\sqrt{3}/8$, $g_{22}=g_{12}=4g_{11}$ and $\eta =2$ in Eqs. (\ref{GPeq1}) and
(\ref{GPeq2}).

Equations (\ref{GPeq1}) and (\ref{GPeq2}) conserve two dynamical invariants,
\textit{viz}., the Hamiltonian and total power (alias the Manley-Rowe
invariant \cite{Luther2000,Porat2013,Phillips2013}):
\begin{eqnarray}
&&H=\int \int \ \left( \mathcal{H}_{k}+\mathcal{H}_{\Omega }+\mathcal{H}_{2}+%
\mathcal{H}_{3}\right) dxdy,  \label{Ham} \\
&&P=\iint {\left( |u_{1}|^{2}+2|u_{2}|^{2}\right) }dxdy\equiv P_{1}+P_{2},
\label{power}
\end{eqnarray}%
where $\mathcal{H}_{k}=\frac{1}{2}|\nabla u_{1}|^{2}+\frac{1}{2\eta }|\nabla
u_{2}|^{2}$, $\mathcal{H}_{\Omega }=-\Omega \left(
|u_{1}|^{2}+|u_{2}|^{2}\right) $, $\mathcal{H}_{2}=\sigma (x)\left(
u_{1}^{\ast 2}u_{2}+\mathrm{c.c.}\right) $ and $\mathcal{H}_{3}=\frac{1}{2}%
g_{11}|u_{1}|^{4}+g_{12}|u_{1}|^{2}|u_{2}|^{2}+\frac{1}{2}g_{22}|u_{2}|^{4}$%
. The power sharing between the FF and SH components is defined as the ratio
$r=P_{1}/P_{2}$. Control parameters for the subsequent analysis are $P$, $%
\ell $ and $\Omega $ (the total power, the stripe's width, and the scaled
detuning).

\section{Results}

\subsection{Numerical results}

Stationary solutions to Eqs. (\ref{GPeq1}) and (\ref{GPeq2}) with
propagation constant $\beta $ were looked for as
\begin{equation}
u_{p}\left( x,y,z\right) =\phi _{p}\left( x,y\right) \mathrm{e}^{ip\beta
z},\quad p=1,2  \label{stationary}
\end{equation}%
where $\phi _{1,2}$ are the stationary amplitudes of the FF and SH
component. Vortex solutions were generated by means of the imaginary-time
(imaginary-$z$) method, applied to Eqs. (\ref{GPeq1}) and (\ref{GPeq2}),
with the input taken at $z=0$ as
\begin{eqnarray}
&&\phi _{1}=r^{|S|}\exp \left( -\alpha _{1}r^{2}+iS\theta \right) ,
\label{ansatz1} \\
&&\phi _{2}=r^{2|S|}\exp \left[ -\alpha _{2}r^{2}+i2S\tilde{\theta}(x)\right]
,  \label{ansatz2}
\end{eqnarray}%
where $r$ and $\theta $ are the 2D polar coordinates, and
\begin{equation}
\tilde{\theta}(x)=\theta +\frac{1}{4S}[\sigma (x)-1]\pi ,
\label{modulationphase}
\end{equation}%
with $S$ and $2S$ representing the winding numbers of FF and SH components,
respectively. In Eq. (\ref{modulationphase}), the matching condition between
the phases of the FF and SH components, $\varphi _{1,2}(x,y)\equiv \arg
\{\phi _{1,2}\}$, is defined by setting
\begin{equation}
\varphi _{2}(x,y)=2\varphi _{1}(x,y)-\varphi _{d}(x,y),  \label{PMC}
\end{equation}%
with $\varphi _{d}=0$ and $\pi $ corresponding, respectively, to $\sigma
(x)=1$ and $-1$ (i.e., $\varphi _{d}=-[\sigma (x)-1]\pi /2$). According to
Eqs. (\ref{ansatz1}) and (\ref{ansatz2}), one has $\varphi _{1}=S\theta $
and $\varphi _{2}=2S\tilde{\theta}(x)$, hence Eq. (\ref{modulationphase}) is
derived via Eq. (\ref{PMC}).

The stability of the stationary vortex solitons was tested by direct real-$z$
simulations of the perturbed evolution in the framework of Eqs. (\ref{GPeq1}%
) and (\ref{GPeq2}) up to $z=10000$. Unstable solutions readily exhibit
splitting in the course of the simulations.

\begin{figure}[th]
\centering
{\includegraphics[scale=0.15]{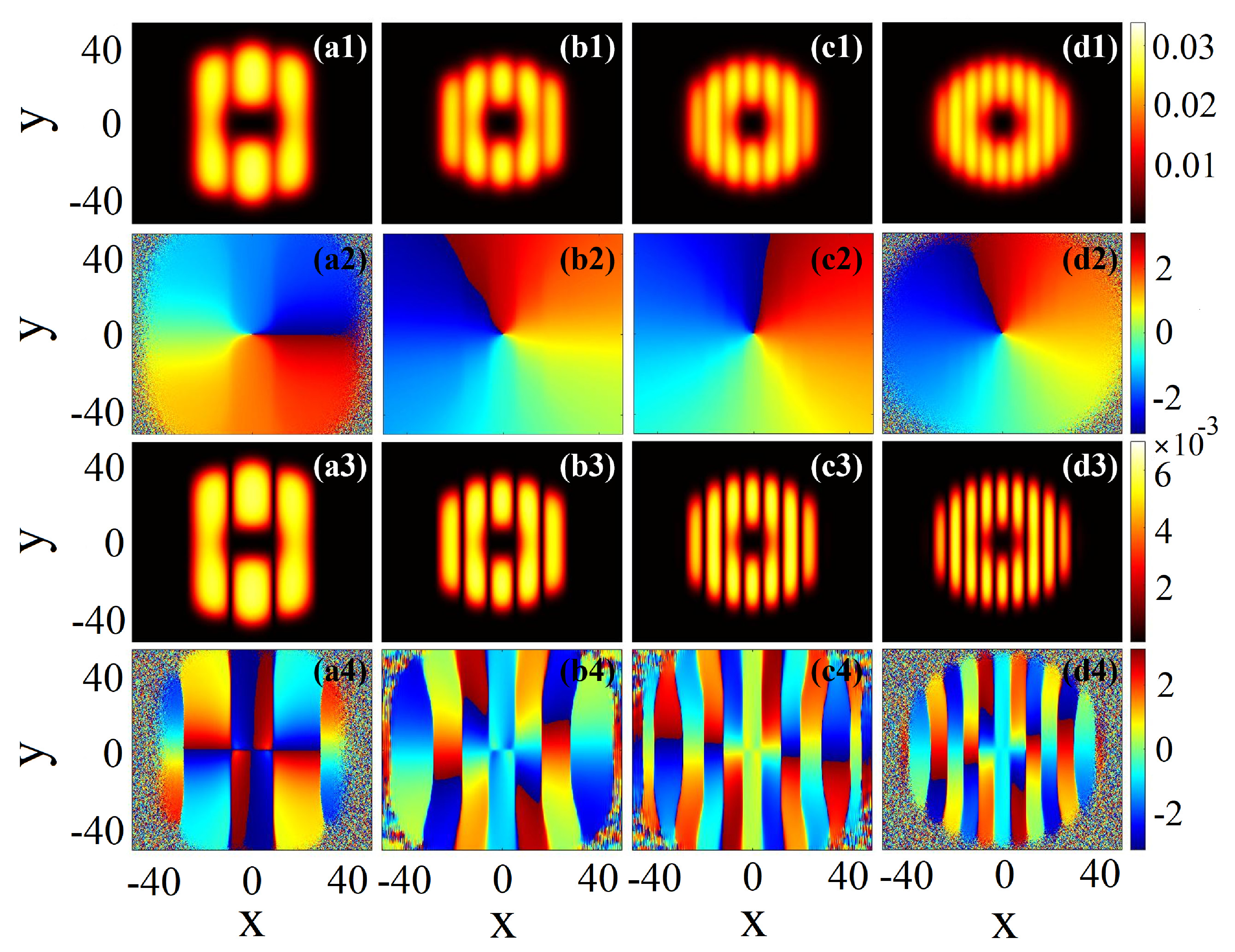}}
\caption{(Color online) Typical examples of the intensity distribution, $|%
\protect\phi _{1}(x,y)|^{2}$ (the first row) and $|\protect\phi %
_{2}(x,y)|^{2}$ (the third row), and phase patterns of $\protect\phi %
_{1}(x,y)$ (the second row) and $\protect\phi _{2}(x,y)$ (the fourth row) of
a semidiscrete vortex optical droplet with $S=1$, which corresponds to point
\textquotedblleft A-D" in the stable area of the OC-type in Fig. \protect\ref%
{stablearea}(a). The parameters are $(P,\ell )=(80,18)$ in (a), $(80,11)$ in
(b), $(80,8)$ in (c), and $(80,6.5)$ in (d). In this case, the effective
detuning is fixed as $\Omega =0$.}
\label{density1}
\end{figure}

\begin{figure}[th]
\centering
{\includegraphics[scale=0.15]{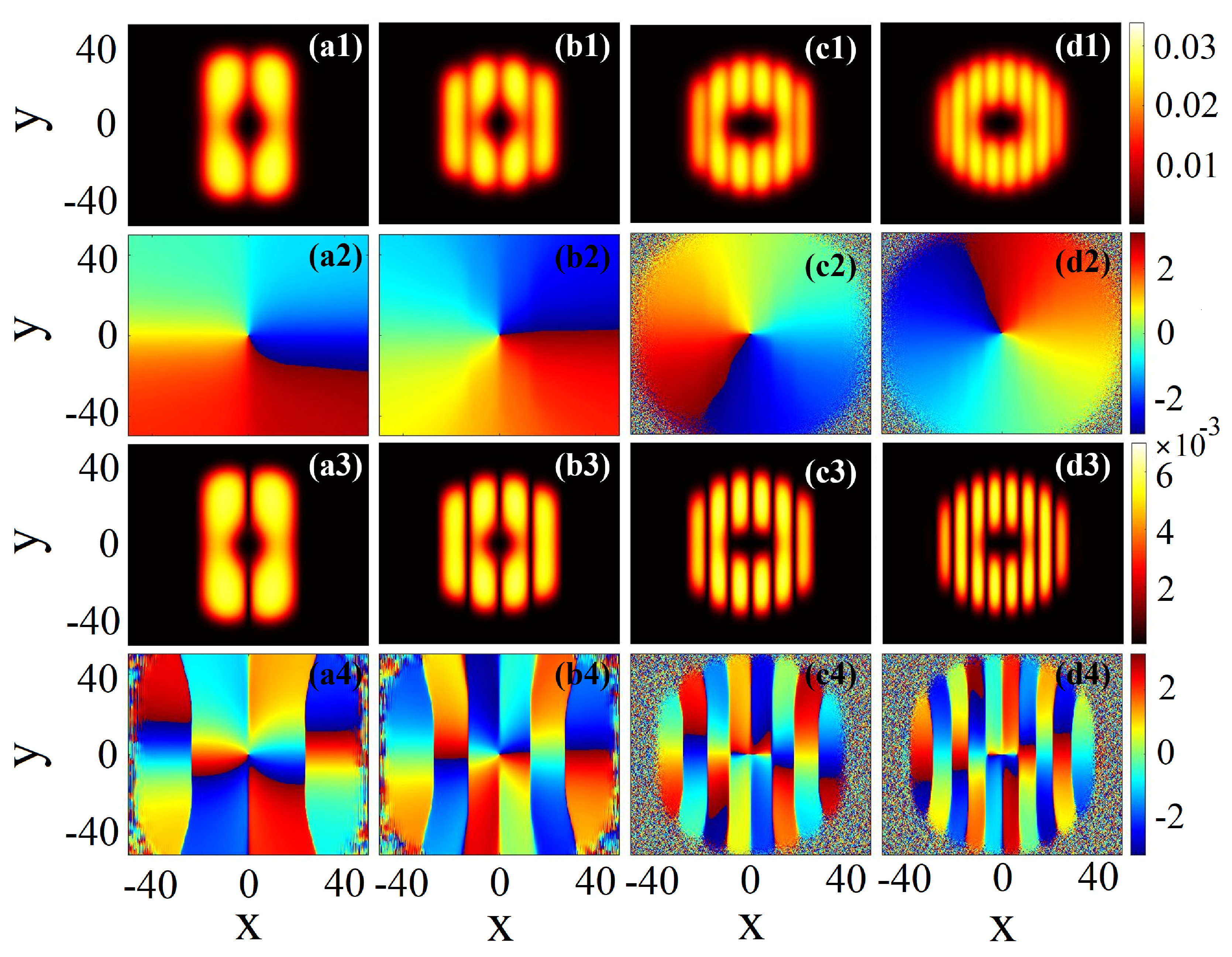}}
\caption{(Color online) A typical example of a stable IC-type optical
droplet with $(P,\Omega )=(80,0)$ and $S=1$. The first and second rows
display the intensity and phase distributions of the FF component, while the
third and fourth rows exhibit the same for the SH component. The stripe's
widths in panels (a-d) are $\ell =22,13,9,$ and $7$, respectively,
corresponding to points \textquotedblleft A-D" in Fig. \protect\ref%
{stablearea}(b).}
\label{density2}
\end{figure}

\begin{figure}[th]
\centering
{\includegraphics[scale=0.6]{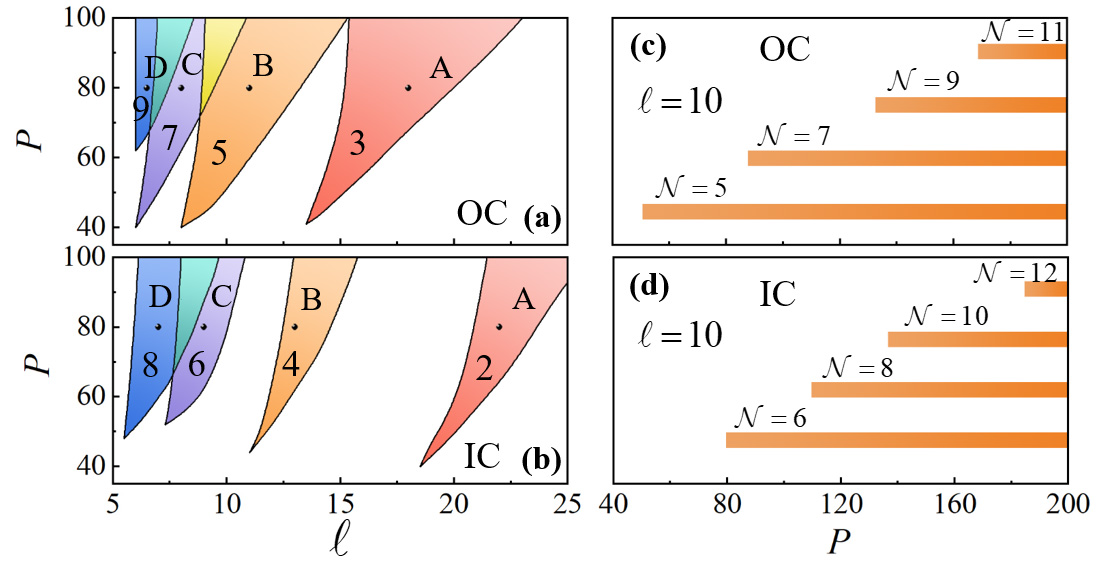}}
\caption{(Color online) Stability areas of semidiscrete vortex optical
droplets of the OC (a) and IC (b) types, in the $(P,\ell )$ plane with $%
\Omega =0$ and $S=1$. Digits in colored stability areas indicate the number
of stripes in the droplets, which are odd in (a) and even in (b). Note that
the presence of bistability in the cyan and yellow areas. Stability ranges
of multistable states are shown in (c) and (d) for the droplets of the OC
and IC types, respectively, with $(\ell ,\Omega )=(10,0)$ and varying values
of the total power, $P$.}
\label{stablearea}
\end{figure}

As the stripe modulation acts solely in the $x$-direction, the vortex
solutions feature similar modulation in the same direction, and can be
characterized by the number of stripes, $\mathcal{N}$, in the localized
solution According to Eq. (\ref{ICOC}), the solutions are also categorized
into the OC and IC types. For the OC- and IC-type solutions, with the pivot
located at the center of a stripe or at the border between adjacent stripes,
numbers $\mathcal{N}$ are, respectively, odd or even. Typical examples for
the OC-type vortex solution with $\mathcal{N}=3,5,7,9$ and IC-type ones with
$\mathcal{N}=2,4,6,8$, all carrying the winding number $S=1$ in their FF
component, are shown, respectively, in Figs. \ref{density1} and \ref%
{density2}. All these states are stable, as confirmed by direct simulations
up to $z=10000$. Because the modulation is applied only along the $x$%
-direction, the vortex solutions feature a typical semidiscrete
configuration similar to that reported in previous works \cite{Zhang2019,Xu2021}. However, unlike those works, the stable vortex
modes are elaborated here in the bulk crystals with the spatially modulated
local $\chi ^{(2)}$ susceptibility. The vortex phase patterns exhibited in
Figs. \ref{density1} and \ref{density2} of the SH component are striped to
obey the matching condition of Eq. (\ref{PMC}), hence, the effective angular
coordinate of this component is defined by Eq. (\ref{modulationphase}),
showing a complicated striped-mixed vorticity pattern of this component.

\begin{figure}[tbph]
\centering
{\includegraphics[scale=0.6]{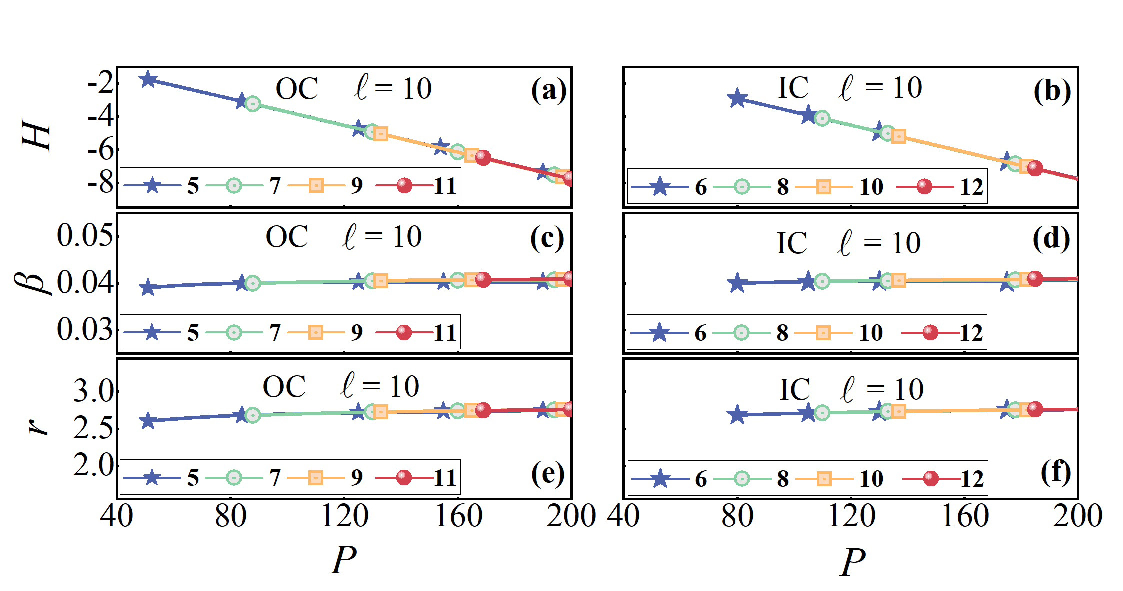}}
\caption{(Color online) (a,b) Hamiltonian $H$, defined as per Eq.(\protect
\ref{Ham}), (c,d) the FF propagation constant $\protect\beta $, and (e,f)
ratio $r=P_{1}/P_{2}$ vs. total power $P$. The lines of blue stars, cyan
circles, orange squares, and red balls in (a,c,e) indicate, respectively,
semidiscrete vortex states with stripe numbers $\mathcal{N}=5,7,9$, and $11$%
, while in (b,d,f), they represent $\mathcal{N}=6,8,10$, and $12$. Other
parameters are $\ell =10,\Omega =0$, and vorticity of fundamental frequency $%
S=1$.}
\label{multistability}
\end{figure}

\begin{figure}[tbph]
\centering
{\includegraphics[scale=0.75]{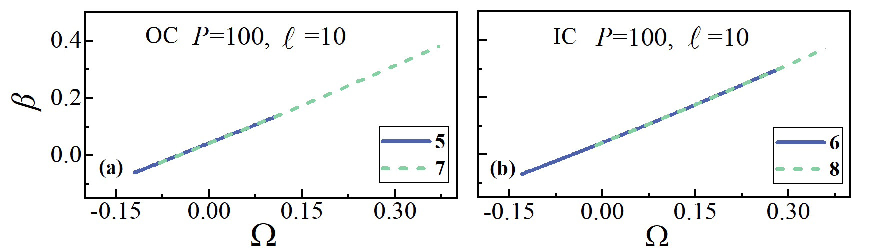}}
\caption{(Color online) (a) and (b): The FF propagation constant $\protect%
\beta $ for stable vortex optical droplets of the OC and IC types vs.
effective detuning $\Omega $. Solid and dashed lines denote the number of
stripes $\mathcal{N}=5$ and $\mathcal{N}=7$ for OC, or $\mathcal{N}=6$ and $%
\mathcal{N}=8$ for IC, respectively. Other parameters are $(P,\ell
)=(100,10) $ and $S=1$.}
\label{mismatch}
\end{figure}

Stability areas for the vortex solutions of the OC and IC types, with
different values of $\mathcal{N}$ are shown, in the $(P,\ell )$ plane with $%
\Omega =0$ and $S=1$, in Fig. \ref{stablearea}(a,b). These plots demonstrate
that the vortex solutions with larger values of $P$ and smaller values of $%
\ell $ produce overlaps between different stability areas [see the cyan and
yellow areas in Fig. \ref{stablearea}(a), and the cyan area in Fig. \ref%
{stablearea}(b)]. The overlaps indicate the presence of multistability in
the system. To further illustrate this feature, we select $\ell =10$ and
examine solutions with different values of $\mathcal{N}$, up to $P=200$. In
this region, we find that four different values of $\mathcal{N}$ stably
coexist at $P=200$ for both OC and IC types of the solutions. Note also that
the solutions with larger values of $\mathcal{N}$ require larger values of $%
P $ to support the stability. Multistable semidiscrete vortex solutions are
characterized by values of $H$, $\beta $ and $r$, which are displayed, as
functions of $P$, in Fig. \ref{multistability}, while keeping $\ell $ and $%
\Omega $ fixed. Notably, curves $H(P)$ and $\beta (P)$ for these solutions
overlap almost completely, indicating degenerate solutions. The nearly flat
dependences, with $d\beta /dP\approx 0$, indicate the existence of broad
states, which may be considered as effectively liquid ones, cf. Ref. \cite%
{Petrov2015}. A nearly constant value $r(P)\approx 2.7$ in Fig. \ref%
{stablearea}(e,f) implies the domination of the FF component in the system.

Finally, in Fig. \ref{mismatch} we present the range of the effective
detuning for a fixed total power, $P=100$. The results indicate that stable
semidiscrete vortex solitons exist for $\Omega \neq 0$.

\begin{figure}[tbph]
\centering
{\includegraphics[scale=0.2]{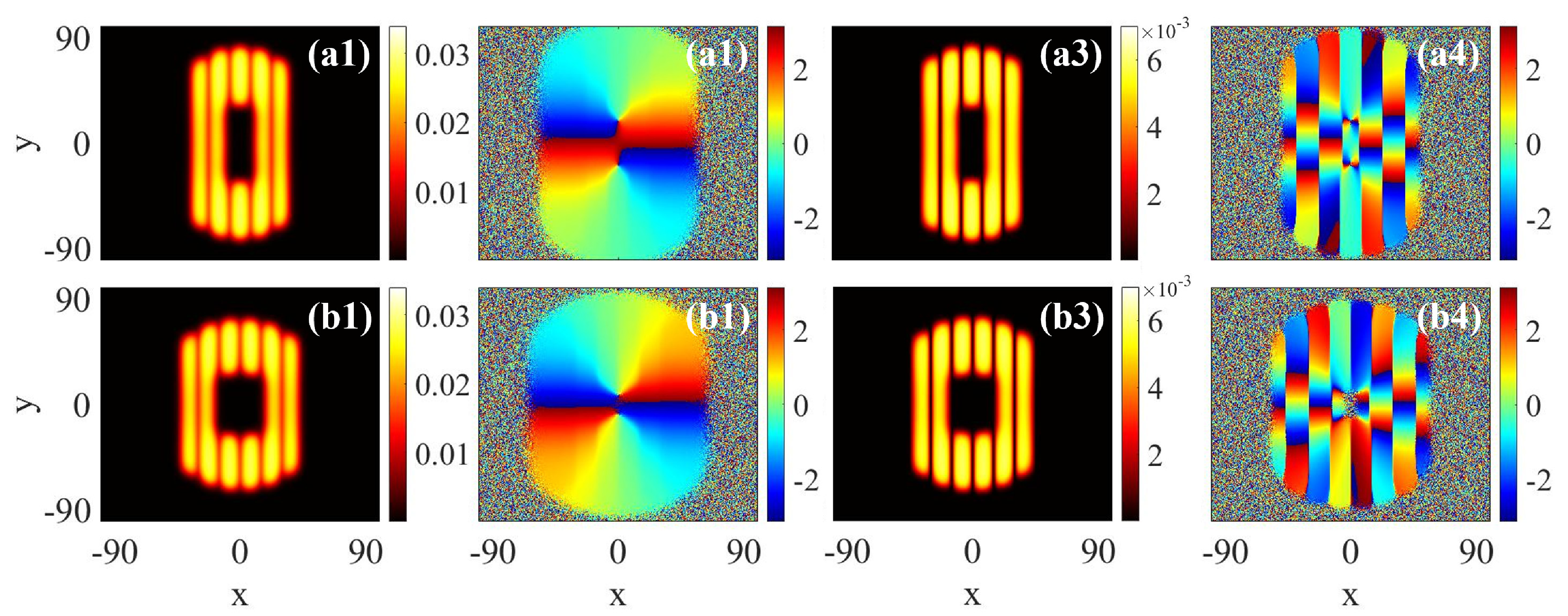}}
\caption{(Color online) (a1)-(a4) A typical example of a stable semidiscrete
vortex droplet of the OC type, with vorticity $S=2$ of the FF component.
Other parameters are $(P,\ell ,\Omega )=(300,15,0)$. (b1)-(b4) An example of
a stable semidiscrete vortex droplet of the IC type, with vorticity $S=2$ of
the FF component, for the same parameters as in (a1)-(a4). The four columns
of panels from left to right display the intensity and phase patterns of the
FF and SH components. These solitons with $S=2$ remain stable for the
propagation distance $z=10000$, up to which the simulations were running.}
\label{S=2}
\end{figure}

Exploring vortex states with $S>1$ is a challenging problem, as they may be
stable only for sufficiently large values of $P$. Two typical examples of
stable vortex solutions of the OC and IC types with $S=2$, and $\mathcal{N}%
=5 $ and $6$, are shown in Fig. \ref{S=2}. The stable solution with $S=2$
exists at $P>280$. This threshold is much higher than its counterpart for $%
S=1$, in which case the stable solutions are found at $P>40$.

\subsection{An outline of the experimental setup}

The fabrication of 3D photonic crystal by means of femtosecond laser pulses
is a mature technology \cite{Keren2018,Arie2021}. To
estimate parameters of the setting under the consideration, we consider the
nonlinear photonic crystal implemented in LiNbO$_{3}$, which has the
second-order nonlinearity coefficient $d_{0}=d_{22}=2.1$ pm/V %
\cite{Dmitriev1991}, and the third-order one $\chi
^{(3)}=6.6\times 10^{-22}$ m$^{2}$/V$^{2}$ \cite{Kulagin2003}. The wavelengths of the FF and SH components are
selected as $1064$ nm and $532$ nm, respectively. The relations between the
scaled units, in which Eqs. (\ref{GPeq1}) and (\ref{GPeq2}) are written, and
their physical counterparts can be established by means of Eq. (\ref{units}%
), as summarizes in Table \ref{data}.
\begin{table}[h]
\caption{Relations between scaled and physical units of the coordinates,
stripe's width, total power, and intensity.}
\label{data}\centering
\begin{ruledtabular}
\begin{tabular}{cc}
\toprule
\midrule
$x=1 \ \& \ y=1$ & 4.64 $ \mu$m \\
$\ell=10\sim15$ & $46.4\sim69.6$ $\mu$m \\
$z=1$ & 280 $\mu$m \\
$P=1$ & 31 kW \\
$|u_{1}|^{2}=0.01 \ \& \ |u_{2}|^{2}=0.01$ & 1.44 GW/cm$^{2}$ \& 0.72 GW/cm$^{2}$ \\
\bottomrule
\end{tabular}
\end{ruledtabular}
\end{table}


According to results of the simulations (see Figs. \ref{density1} and \ref%
{density2}), the peak droplet's intensity in the FF and SH components are $%
I_{\mathrm{FF}}\approx 4.32$ GW/cm$^{2}$ and $I_{\mathrm{SH}}\approx 0.432$
GW/cm$^{2}$, respectively. If we select the pulse width as $200$ ps, the
energy densities for the FF and SH components are $0.86$ J/cm$^{2}$ and $%
0.086$ J/cm$^{2}$, which are lower than the damage threshold ($\sim $1.5 J/cm%
$^{2}$ \cite{damage1,damage2}) of the PPLN crystals. The characteristic
propagation distance, which is $z=10000$, amounts to $2.8$ m, which is
several times the underlying diffraction length. Therefore, the stability of
the solitons, predicted by the simulations, is a reliable prediction.

The sketch for the experimental observation of these droplets is shown in
Fig. \ref{stripefig}(d): A high-power sub-nanosecond laser (e.g., one with
the power exceeding $0.6$ mJ per pulse, $200$ ps pulse width, and $400$ Hz
repetition rate) may be a suggested light source in the proposed experiment.
The input patterns on the front surface of the PPLN can be generated by a
spatial light modulator (SLM) and a Lens. Input FF and SF components can be
selected with energies $0.37$ mJ and $0.16$ mJ per pulse, which are closed
to the FF/SF power ratio $r\approx 2.6$ for stationary semidiscrete vortex
droplets in Figs. \ref{mismatch}(e) and (f). If the input's power ratio is
essentially different from this value, it naturally gives rise to strong
oscillations between the FF and SH components. The power and phase patterns
of the FF component of the input, which are shown in the left inset of Fig. %
\ref{stripefig}(d), can be produced by a properly designed SLM1 and L1, and
the input SH patterns with $0.14$ mJ per pulse can be produced by SLM2 and
L2, respectively. The FF and SH beams are coupled by the dichroic mirror,
which sends them onto the PPLN coaxially. The necessary PPLN crystal may be $%
4$ cm long along the axial direction. Finally, the beams transmitted through
the PPLN are imaged on to a camera by Lens L3. The simulated output patterns
at the back surface of the crystals are shown in the right inset in Fig. \ref%
{stripefig}(d).

\section{Conclusion}

We have proposed 3D photonic crystals with the striped structure and the
combination of the $\chi ^{(2)}$ and defocusing $\chi ^{(3)}$
nonlinearities. The results of the analysis predict the creation of two
types of stable semidiscrete vortex solutions, OC and IC (onsite- and
intersite-centered ones), which exhibit an odd or even number of stripes in
their structure, respectively. The smallest number of the stripes is $%
\mathcal{N}_{\text{OC}}=3$ or $\mathcal{N}_{\text{IC}}=2$. The setting
admits multistability, \textit{viz}., the coexistence of stable solutions
with different numbers of stripes for the same parameters. Unlike the
multistability in systems with the competing cubic-quintic nonlinear
interactions, the Hamiltonian and propagation constant of these coexisting
states are equal, thus featuring degeneracy. The range of the multistability
has been found. The stable solutions for semidiscrete vortices exist within
a certain range of positive or negative values of the phase mismatch of the $%
\chi ^{(2)}$ interaction in the medium.

The scheme proposed in this paper can be developed in other settings, such
as photonic crystals with ring- or fan-shaped structures. Those settings can
be used, in particular, for implementation of various scenarios of beam
shaping.

\begin{acknowledgements}
This work was supported by the NNSFC (China) through Grants No. 12274077,
11874112, by the Research Fund of Guangdong-Hong Kong-Macao Joint Laboratory for Intelligent Micro-Nano Optoelectronic Technology through
grant No.2020B1212030010 and the Graduate Innovative Talents Training
Program of the Foshan University. The work of B.A.M. is supported, in part,
by the Israel Science Foundation through grant No. 1695/22.
\end{acknowledgements}

\end{document}